\documentstyle[11pt,aaspp4]{article}

\lefthead{Watanabe et al.}
\righthead{The Diffuse Gamma-Ray Background from Supernovae}

\begin{document}

\title{The Diffuse Gamma-Ray Background from Supernovae}

\author{K. Watanabe}
\affil{USRA/LHEA, NASA/Goddard Space Flight Center, \\
Code 660.2, Greenbelt, MD 20771}

\author{D. H. Hartmann, M. D. Leising, and L. -S. The}
\affil{Department of Physics and Astronomy, \\
Clemson University, Clemson, SC 29634-1911}

\begin{abstract}

The diffuse extragalactic $\gamma$-ray background in the MeV region is believed
to be due to photons from radioactivity produced in supernovae throughout the 
history of galaxies in the universe. In particular, $\gamma$-ray line emission 
from the decay chain \(^{56}{\rm Ni} \rightarrow \)  \(^{56}{\rm Co} 
\rightarrow\) \(^{56}{\rm Fe} \) provides the dominant photon source 
(\cite{CLAY69}). Although iron synthesis occurs in all types of supernovae, 
the contribution to the background is dominated by Type Ia events due to their 
higher photon escape probabilities. Estimates of the star formation history in 
the universe suggest a rapid increase by a factor $\sim$ 10 from the present 
to a redshift z$_p$ $\sim$ 1.5, beyond which it either remains constant or 
decreases slowly. Little is known about the cosmological star formation 
history for redshift exceeding z $\sim$ 5. We integrate the observed star 
formation history to determine the Cosmic Gamma-Ray Background (CGB) from the 
corresponding supernova rate history. In addition to $\gamma$-rays from 
short-lived radioactivity in SNIa and SNII/Ib/Ic we also calculate the minor 
contributions from long-lived radioactivities ($^{26}${\rm Al}, $^{44}${\rm Ti},
$^{60}${\rm Co}, and electron-positron pair annihilation). The time-integrated 
$\gamma$-ray spectrum of model W10HMM (\cite{PINT88A}, \cite{PINT88B}) was used
as a template for Type II supernovae, and for SNIa we employ model W7 
(\cite{NOMO84}). Although progenitor evolution for Type Ia supernovae is not 
yet fully understood, various arguments suggest delays of order 1$-$2 Gy 
between star formation and the production of SNIa's. The effect of this delay 
on the CGB is discussed. We emphasize the value of $\gamma$-ray observations of
the CGB in the {\rm MeV} range as an independent tool for studies of the 
cosmic star formation history. If the delay between star formation and SNIa 
activity exceeds 1 Gy substantially, and/or the peak of the cosmic star 
formation rate occurs at a redshift much larger than unity, the $\gamma$-ray 
production of SNIa would be insufficient to explain the observed CGB and a so 
far undiscovered source population would be implied. Alternatively, the cosmic 
star formation rate would have to be higher (by a factor 2-3) than commonly
assumed, which is in accord with several upward revisions reported in the recent
literature. 
\end{abstract}


\section{Introduction}

The study of extragalactic background radiation in various wavelength bands 
holds the keys to many important astrophysical questions. While the cosmic 
microwave background is an imprint of conditions in the early universe, the 
high-energy background contains information on galaxy evolution, stellar 
explosions, and processes near supermassive black holes in the cores of active 
nuclei. Here we focus on the energy window between 100 keV and 10 MeV, an 
observationally difficult region for which recent analyses of data from 
COMPTEL (Kappadath {\it et al.} 1996), HEAO-1 A4 (Kinzer {\it et al.} 1997), 
and SMM (Watanabe {\it et al.} 1997) now provide reasonably accurate 
measurements. Below a few hundred keV the observed Cosmic Gamma-Ray Background
(CGB) is believed to be due to the superposition of unresolved Seyfert 
galaxies (e.g., Zdziarski 1996),  while for photon energies above 
3 MeV blazars are the dominant source population (e.g., Sreekumar, Stecker, 
$\&$ Kappadath 1997). There is no known galaxy type that can fill the gap 
between the Seyfert galaxies and the blazars. However, the CGB around photon 
energies \( E_{\gamma} \sim 1 {\rm MeV} \) could be partly, or even completely,
due to cumulative $\gamma$-ray production in supernovae. In particular, 
$\gamma$-ray lines from the reaction chain \( ^{56}{\rm Ni} \rightarrow \) 
\(^{56}{\rm Co} \rightarrow \) \(^{56}{\rm Fe} \) 
are abundant enough to generate a detectable signal 
(\cite{CLAY69}; \cite{CLAY75}; \cite{THE93}). 

Although iron synthesis occurs in all SN types, the supernova $\gamma$-ray 
contribution to the CGB is dominated by $\gamma$-ray escape from the iron 
synthesis in Type Ia supernovae (SNI). This can be understood from 
considerations of their rates (R), yields (M, ejected $^{56}$Ni mass), and 
likelihood of photon escape. The rate-yield product satisfies
${\rm R_{I}\ M_{I} \sim R_{II}\ M_{II}}$, because 
${\rm R_{I} \leq 0.2 R_{II}}$, ${\rm M_{I}
\sim 0.5 M_{\odot}}$, and ${\rm M_{II} \sim 0.1 M_{\odot}}$. 
Here we use the label II to represent core collapse events, lumping together 
Type II and Type Ib/c supernovae. The reason that Ia's dominate the CGB is 
thus ultimately due to the fact that their photon escape probablilities are 
much higher than those for SNII. To determine the SN-induced CGB it is not 
necessary to use SN light curves, but one must distinguish between radioactive 
nuclei with lifetimes short enough for $\gamma$-ray transport to occur in a 
opaque or semi-opaque expanding atmosphere, for which time-integrated light 
curves are needed, and radioactive nuclei whose $\gamma$ rays are produced long 
after the supernova remnant has become transparent. In $\S$2 we introduce the 
basic formalism using monochromatic lines from long-lived isotopes. In $\S$4 
we discuss the modifications required to treat $\gamma$-ray transport in cases 
of isotopes with short lifetimes. We discuss SNI and SNII separately, and 
emphasize the need to incorporate the delay between star formation activity and
its associated SNI activity. We then compare the CGB estimates to the 
observations. 

\section{Basic Formalism}

The CGB from supernovae is due to a mixture of $\gamma$-rays from various
radioactivities. Short-lived isotopes release photons that are still affected 
by the expanding supernova envelope, and detailed transport calculations are 
required to determine the emerging spectrum. This strongly affects photons 
from the decay of $^{56}$Ni, while long-lived isotopes such as $^{26}$Al and 
$^{44}$Ti release photons into an optically thin medium. We begin the 
discussion of the formalism by first considering long-lived nuclei. The 
ejection of some mass ($M_{ej}$) of radioactive material then generates, 
assuming 1$\gamma$ per decay, a total photon number

\begin{eqnarray}
N_{\gamma} = \frac{M_{ej}}{A\ u} = 1.2 \times 10^{53}
A^{-1} M_{-4},
\end{eqnarray}
where $A$ is atomic mass number of the radioactive element, u is
the atomic mass unit, and
M$_{-4}$ = M$_{\rm ej}$ / 10$^{-4}$ M$_{\odot}$ is a normalized ejecta mass.

The production rate of photons is proportional to the supernova rate, which we 
take to be proportional to the star formation rate
\begin{eqnarray}
 R_{SN} = \xi_{SN} \dot{\rm M}_{\star}
\end{eqnarray}
where $\dot{\rm M}_{\star}$ is the star formation rate in 
M$_{\odot}{\rm y}^{-1}$, and $\xi_{SN}$ is measured in units of 
M$_{\odot}^{-1}$. Estimates of the present-day star formation rate in the 
Milky Way vary significantly, but within a factor two one finds
$\dot{\rm M}_{\star} \sim {\rm few}\  {\rm M_{\odot}} {\rm y}^{-1}$ and R$_{SN}$
$\sim$ few 10$^{-2}$ y$^{-1}$ (e.g., \cite{TIMM97}). Therefore, we expect that
$\xi_{SN}$ is of order 10$^{-2}$ M$_{\odot}^{-1}$. We assume that this 
normalization can be universally applied to star formation, regardless of 
galaxy type and whether or not a given galaxy undergoes quiescent or bursting 
star formation. We thus simply apply the scaling relation to the global star 
formation rate density. Recent estimates of this density in the local 
universe using H$_{\alpha}$ observations (\cite{COLE94}; \cite{GALL95}) yield 
the global star formation rate density,  

\begin{eqnarray}
\dot{\rho}_{\star}(z=0) = \dot{\rho}_{\star}(0) \simeq 3.7 \times
10^{-2}\ {\rm h}^{3}\ {\rm M_{\odot}}\  {\rm Mpc}^{-3}\  {\rm y}^{-1},
\end{eqnarray}
where the Hubble constant is
H$_{0}$ = 100\ h \ km\ s$^{-1}$\ Mpc$^{-1}$. This value is rather similar to
earlier estimates based on colors of field galaxies (\cite{TD80}).  
The corresponding local $\gamma$-ray emissivity 
(measured in $\gamma {\rm Mpc}^{-3} {\rm y}^{-1}$) is given by
\begin{eqnarray}
 \dot{n}_{\gamma}(0) = \dot{\rho}_{\star}(0)\  \xi_{SN} \ N_{\gamma}.
\end{eqnarray}
Here we ignore the fact that on scales less than 300 Mpc (redshift less than z 
= 0.1) the matter distribution, and thus also the star formation distribution, 
is non-uniform. We comment on this point briefly in the Conclusions. The 
$\gamma$-ray production is treated as a uniform source density, which evolves 
with redshift in accordance with the evolving star formation rate. 

The emissivity is then integrated over distance (redshift) to
yield the differential CGB flux 
\begin{eqnarray}
 \frac{\partial F}{\partial \Omega} (\gamma\ cm^{-2} s^{-1} sr^{-1})
= F_{0} {\em C},
\end{eqnarray}
where C is a dimensionless combination of cosmological factors, and with the 
usual definition of Hubble length, L$_{\rm H}$ = c / H$_0$, the normalization 
is given as
\begin{eqnarray}
 F_{0} = (4\pi)^{-1} \dot{n}_{\gamma}(0)\ L_{H} \simeq 3.53
\times 10^{-3} h^{2} A^{-1} M_{-4} \xi_{SN} \ . 
\end{eqnarray}
The main integral to be performed is  
\begin{eqnarray}
 C = \int^{\infty}_{0} dz (1+z)^{-1} E(z)^{-1}\eta(z) \ ,
\end{eqnarray}
where $\eta(z)$ contains evolutionary effects to be discussed below, the 
factor (1+z)$^{-1}$ accounts for the dilation of the supernova rate, and 
the function $E(z)$ (equation (13.3) in \cite{PEEB93}) 
\begin{eqnarray}
 E(z) = \left[ \Omega(1+z)^{3} + \Omega_{R}(1+z)^2 +
\Omega_{\Lambda} \right]^{1/2},
\end{eqnarray}
represents the evolution of the Hubble ``constant"
\begin{eqnarray}
\frac{\dot{a}}{a} = H_{0}E(z) = H(z) \ ,
\end{eqnarray}
with $a$(t) being the scale factor of the universe.
To recover the present-day Hubble constant H$_0$ it is required that
\begin{eqnarray}
\Omega + \Omega_{R} + \Omega_{\Lambda} =1 \ ,
\end{eqnarray}
where the terms represent the densities of matter, curvature, and a cosmological
constant relative to the critical density of the universe. 
For \( \Omega_{\Lambda} = 0 \), the E(z) function simplifies significantly 
\begin{eqnarray}
 E(z) = (1+z)(1+\Omega z)^{1/2}.
\end{eqnarray}
We assume $\Lambda$ = 0, h = 0.75, and $\Omega$ = 1
for the cosmological parameters of our standard model, motivated by the 
remarkable match between theory and observation of the cosmic microwave 
background and large scale structure power spectrum (e.g., \cite{GS98} and 
references therein).

The SN-induced CGB depends sensitively on the evolution of the cosmic star 
formation rate. Recent progress in this field now provides measurements 
of this evolution to redshift well past z = 1. We already introduced (eq. 3) 
the local rate density derived from H$_\alpha$ surveys of nearby galaxies 
(e.g., \cite{GALL95}) to which we normalize the star formation history. 
Conversion of the H$_\alpha$ luminosity function of galaxies in the local 
universe can be related to the star formation rate under the assumption of an 
IMF. The value from \cite{GALL95}, which we employ in this study, is based on 
the Salpeter IMF. Recent UV observations of \cite{TREY97} and 
H$_\alpha$ data of \cite{TRES98} suggest 
a factor two increase by redshift z $\sim$ 0.2, which suggests $\alpha$ $\sim$ 
3$-$4 for an evolutionary law of the form SFR(z) $\propto$ (1+z)$^\alpha$, i.e.,
a rather rapid increase of the SFR with redshift. 

How much do we know about the star formation rate density as a function of 
redshift? Besides chemical evolution evidence for a significant increase in the
star formation rate of the Milky Way as a function of look-back time, similar
evidence for enhanced past star formation was found for faint galaxies at 
redshift beyond z $\sim$ 0.3 (\cite{ELL96}; \cite{COW97}; \cite{LILL96}). A 
strong increase in the comoving star formation rate density has also been 
predicted by cosmic chemical evolution models (e.g., \cite{PEI95}) addressing 
the observations of Lyman absorption systems in QSO spectra. The advent of the 
Hubble Deep Field, HDF (\cite{WILL96}) allowed photometric surveys to probe to 
z $\sim$ 5, and \cite{MADA96} showed that the SFR appears to peak around z = 
1.5, and further suggested that the SFR slowly declines to present-day values 
by redshift z $\sim$ 5. Since then, analysis of individual galaxies in fields 
flanking the HDF (\cite{Guzman98}) has supported the rapid increase in the SFR 
to redshift near unity, and combined HST and ground-based IR photometry of the 
HDF (\cite{CONN97}) confirmed that the SFR increases by at least an order of 
magnitude to z $\sim$ 1 and that a peak in the comoving rate density occurs 
between z $\sim$ 1 and z $\sim$ 2. 

Although the detailed shape of the function describing the cosmic star formation
history has not yet been determined, especially its behavior at large redshift,
at least one robust and consistent result appears to emerge from the data: The 
comoving star formation rate density has decreased by about one order of 
magnitude since its peak at $z_p \sim 1 - 1.5$ (\cite{Guzman98}). We thus 
represent the normalized SFR function by a  simple function (see 
\cite{YUNG98} for a similar approach)
\begin{eqnarray}
 \log(\eta(z)) = \left\{ \begin{array}{ll}
                           {\rm A}\ \log(1+z) & \mbox{for $z \leq z_{p}$} \\
{\rm A}\ \log(1+z_{p}) - {\rm B}\ (z-z_{p}) & \mbox{for $z > z_{p}$}
                                 \end{array}     \right.,
\end{eqnarray}
where we select a ``standard'' case (A, B, $z_{p}$) = (4, 0.25, 1) for 
presentation in the figures. For fixed B = 0.25, we explore variations in A and
$z_p$ to constrain the location of the peak of the star formation history with 
the CGB, and we also consider cases with constant star formation rate before 
the peak (B = 0) in order to address the possibility that some significant 
fraction of the star formation activity in the early universe could be hidden 
by dust (e.g., \cite{HUGH98}). In all cases star formation is assumed to be 
zero beyond z = 5. As far as the nuclear yields are concerned we apply 
standard supernova models, neglecting corrections due to the changing metal 
content in the host galaxies. These effects are expected to be of second order,
but might be worthy of further study as our understanding of the CGB improves. 
Supernova surveys suggest a separate treatment of SNI and SNII, and a delay 
between these two event classes. We discuss this point below in more detail. 

The distribution of events with redshift broadens the monochromatic 
$\gamma$-ray lines into an observable continuum. We take this into account 
following standard prescriptions (\cite{WEIN72}, \cite{PEEB93}) and calculate 
the differential flux per unit energy.
 
\section{Long-Lived Isotopes}

For isotopes that decay with a lifetime exceeding a few years the expanding 
atmosphere no longer provides enough optical depth to alter the line spectrum. 
The emerging lines are thus treated with the formalism described above, 
assuming that all photons have a 100$\%$ escape probability. We consider 
$^{26}$Al, $^{44}$Ti, $^{60}$Co, and positrons. 
 
$^{26}$Al has a half-life of \( 7.2 \times 10^{5}\) years, and predominantly 
decays into an excited state of $^{26}$Mg via $\beta ^{+}$-decay and electron 
capture. A single $\gamma$ ray at 1.8 MeV is emitted (See Fig. 3.6 of 
\cite{ARNE96}). We assume that each Type II supernova ejects 
\(M_{26} = 1.0 \times 10^{-4}M_{\odot} \) of radioactive aluminum 
(\cite{TIMM95}). 

$^{44}$Ti is believed to be the dominant nucleosynthetic progenitor of stable
$^{44}$Ca (\cite{BCF68}; \cite{ARNE96}). $^{44}$Ti decays (half-life of $\sim$ 
60 years) to $^{44}$Sc. The decay of $^{44}$Sc (half-life of 3.93 hours) to 
$^{44}$Ca generates a $\gamma$-ray photon (98.99 \% of the time) at E = 1.157
MeV. We assume a typical supernova to eject \(M_{44} = 5.0 \times 
10^{-5}M_{\odot} \) (\cite{TIMM96}). $^{60}$Co decays with a half-life of 5.3 
years to $^{60}$Fe, emitting two $\gamma$-ray photons at E = 1.17 MeV and 
E = 1.33 MeV. A characteristic ejecta mass ($M_{60}$) is $3 \times 10^{-5} 
M_{\odot}$ (\cite{TIMM96}). 

The diffuse 511 keV glow of the Galaxy (e.g. \cite{PRAN93}) is assumed to be 
similar to that other galaxies. We include this line by scaling it to the SNI
rate, assuming that 3\% of all $^{56}$Co positrons escape from SNI and find 
their way into the ISM. The resulting cosmological 511 keV feature is shown in
Figure~\ref{cgbsnii}. Such a fraction would explain most of the Galactic 
annihilation line (e.g. \cite{PURC97}). However, this line will always be 
dwarfed by the contribution from annihilation of $^{56}$Co positrons in the 
supernova ejecta. Averaged over the supernova event, roughly 40\% of the 511 
keV photons from the 97\% of the positrons that annihilate in the SNI 
envelopes escape the ejecta, which will necessarily exceed the diffuse 
emission corresponding to 100\% of the photons from 3\% of the positrons.

\section{Isotopes with short lifetimes}

For isotopes with short half-lives, Compton scattering in the expanding 
supernova is important and detailed $\gamma$-ray transport studies are 
required to determine the emerging spectrum. The formalism presented above only
needs a small modification, through the introduction of the emerging 
differential spectrum normalized to the number of primary $^{56}$Ni nuclei. 

\subsection{Type II supernovae}

\begin{figure}
\plotfiddle{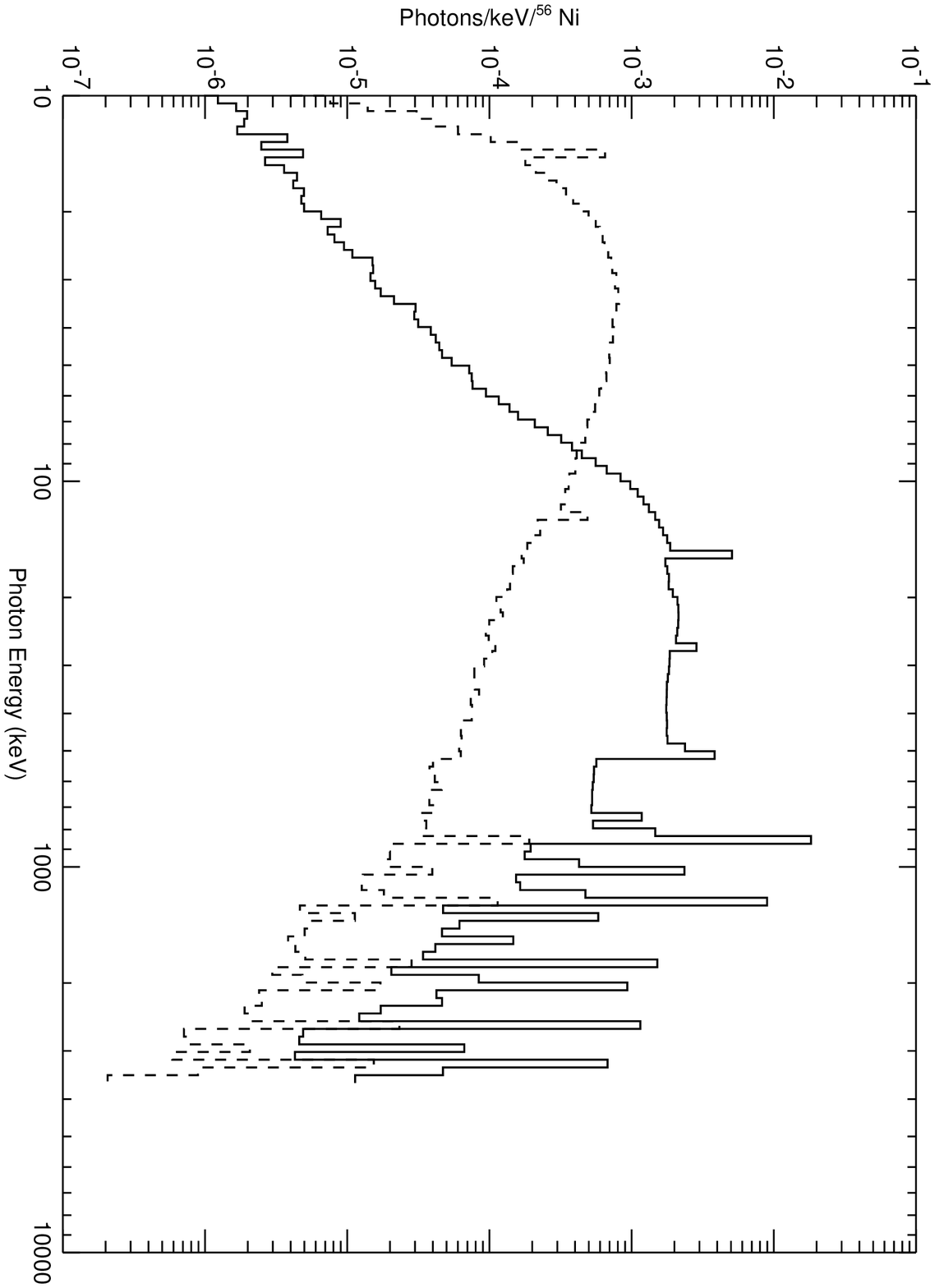}{300pt}{90}{80}{80}{324}{1}
\caption{The time integrated $\gamma$-ray continuum flux per $^{56}$Ni nucleus 
for the SNII template model W10HMM (dashed line), and for the SNI template
model W7fm (solid line). SNI dominate in the MeV regime, because of their 
larger fraction of escaping $\gamma$-rays. SNII dominate at lower energies, 
but in this regime the CGB is overwhelmingly determined by X-ray emission from 
Seyfert galaxies. 
\label{IaII}}
\end{figure}

The light curve of Type II supernovae (SNII) is partially powered by the energy
deposition from the decay chain  \(^{56}{\rm Ni} \rightarrow \)  \(^{56}{\rm 
Co} \rightarrow\) \(^{56}{\rm Fe} \). Some of the photons emitted in this 
process escape and thus contribute to the CGB. $^{56}$Co ( \(t_{1/2} = 77.12d 
\)) decays to $^{56}$Fe, emitting $\gamma$-rays at 0.847 MeV (100\%), 1.04 MeV 
(14\%),1.24 MeV (68\%), 1.77 MeV (16 \%), 2.03 MeV (12 \%),2.6 MeV (17\%), and 
3.24 MeV (12.5\%) among other lines (See Fig. 13.4 of \cite{ARNE96} for a 
{\it simplified} decay scheme). Compton scatterings degrade these photon 
energies, leading to a broad continuum spectrum that develops underneath the 
line spectrum. A typical SNII line photon has a small escape probability of 
order 1$\%$, because of the massive hydrogen envelope. Our treatment of photon 
transport in an expanding supernova is described in \cite{THE90}. As standard 
input for the expanding envelope we use model W10HMM (\cite{PINT88B}), time 
integrated, to provide a SNII template for $\gamma$-ray photons resulting from 
$^{56}$Ni decay. This source function, S(E) - shown in (Figure~\ref{IaII}), is 
used to calculate the CGB flux from the ``prompt'' continuum escaping from SNII.
The number of $\gamma$-rays per 
unit energy as a function of energy is thus given by
\begin{eqnarray}
 N_{\gamma}(E) = N_{56}\ S(E)\  \ \ {\rm (\gamma/keV)}  \ .
\end{eqnarray}
The differential CGB flux is 
\begin{eqnarray}
 \frac{\partial ^{2} F(E_{\gamma})}{\partial E {\em \partial \Omega}}
 ({\em \gamma} {\em cm^{-2}} {\em keV^{-1}} {\em s^{-1}} {\em sr^{-1}}) =
 F_{0} {\em C_{con}} \ \ ,
\end{eqnarray}
where F$_0$ is the same as before, but the integration over the cosmic star
formation history is modified to
\begin{eqnarray}
 C_{con} = \int^{\infty}_{0} dz\ E(z)^{-1}\ \eta(z)\
{\rm S}(E_{\gamma} \times (1+z)).
\end{eqnarray}
The factor (1+z)$^{-1}$ in the previous C function is now absent, because of a 
compensating factor (1+z) due to the compression of the bandwidth in photon 
energy.  

To estimate the parameter $\xi_{SN}$ that appears in the normalization $F_0$ we 
assume a Salpeter IMF in a mass range between 0.1 M$_\odot$ and 125 
M$_\odot$. If Type II supernovae result from stars with masses exceeding $\sim$
8 M$_\odot$, one finds $\xi_{II}$ = 0.007 (see eq. 3 in \cite{MADA98}), which is
in agreement with arguments based on average Galactic chemical evolution 
(\cite{TIMM97}). For the ejected nickel mass in SNII we assume a 
value of \( M_{-4} = 7.5 \times 10^{2} \) (\cite{PINT88B}).

\placefigure{IaII}

\subsection{Type Ia Supernovae}

We now estimate the contribution from SNI, which turn out 
to be the dominant contributors to the CGB in the MeV window.
The method of calculation for the emerging continuum spectrum is identical to 
that used for SNII. As a standard model we use the fully mixed version of W7 
(\cite{NOMO84}), integrated over 600 days (see Figure~\ref{IaII} ). The figure 
shows the reason why SNI contributions to the CGB dominate: the escape fraction
in the MeV regime is more than an order of magnitude larger than that of SNII. 
The normalization in terms of the ejected $^{56}$Ni mass is \( M_{-4} = 5 
\times 10^{3}\), i.e., 0.5 solar masses of iron is produced in an average SNI. 
This should not be considered a rigorously uniform mass, since despite the 
impressive uniformity of SNI light curves it is well recognized that SNI in 
fact do show a significant spread in light curves, peak brightness, and 
spectral evolution. Some of this intrinsic spread is most likely due to 
different nickel masses. The observation of sub-luminous SN 1991bg 
(\cite{TURR96}; \cite{MAZZ97}) suggested a $^{56}$Ni mass of 0.07 M$_\odot$ 
(similar to values commonly attributed to SNII), while SN 1994D apparently 
required between 0.5 and 1.0 M$_\odot$ (\cite{VL96}). However, we are only 
interested in the mean ejecta, because the CGB data have no bearing on any 
individual event.  

While there is essentially no time delay between star formation and subsequent 
SNII because of the short main sequence lifetime of massive stars, a 
significant delay may have to be included for SNI. Chemical evolution studies 
of the Milky Way have often been used to argue for a delay of more than 1 Gy, 
based on a break in the [O/Fe] vs. [Fe/H] distribution of stellar abundances 
(see \cite{PAGE95} and \cite{YOSH96} for recent discussions). Population 
synthesis models using various SNI progenitor schemes suggest a range of 
possible delays from zero to several Gy (\cite{RUIZ98}; \cite{YUNG98}; 
\cite{SADA98}; \cite{MADA98}). A measurement of the delay time scale would go a
long way towards understanding or constraining SNI progenitor models. Such a 
measurement could be achieved with deep SN surveys, measuring the comoving I/II
ratio as a function of redshift (e.g., \cite{RUIZ98}; \cite{SADA98}). While 
supernova searches now routinely detect SNI at z $\sim$ 1 (\cite{PERL97}, 
\cite{TONR97}) and the comoving SNI rate at z $\sim$ 0.4 has already been 
determined by the Supernova Cosmology Project (\cite{PAIN97}), a similar
accomplishment for SNII will have to wait for the {\it Next Generation Space 
Telescope}, NGST (\cite{MADA98}). The present-day average SNI fraction of the 
total rate is $\sim$ 20$\%$, and the I/II ratio is $\sim$ 1/3 (e.g., 
\cite{CAPP97}), but varies with redshift due to the delay between SNI and and 
SNII (see Fig. 5a in \cite{YUNG98}). Depending on the assumed delay time 
between SNI and SNII we adjust $\xi_{I}$ such that the I/II ratio at z = 0 is 
fixed at the observed value 1/3. This prescription fixes the SNI rate 
evolution as a function of redshift in terms of two constants (delay time, and 
I/II ratio), but is otherwise determined by the global star formation history 
and its associated SNII rate (fixed through $\xi_{I}$). This prescription is 
convenient, but not necessarily realistic. However, the lack of rigorous 
theoretical or observational guidance about the evolution of the I/II ratio 
leaves us little choice. Continued efforts to determine supernova rates as a 
function of redshift will eventually lead to a better model, which would 
further improve estimates of the supernova contribution to the CGB in the MeV 
regime.  

\section{Results}

\begin{figure}
\plotfiddle{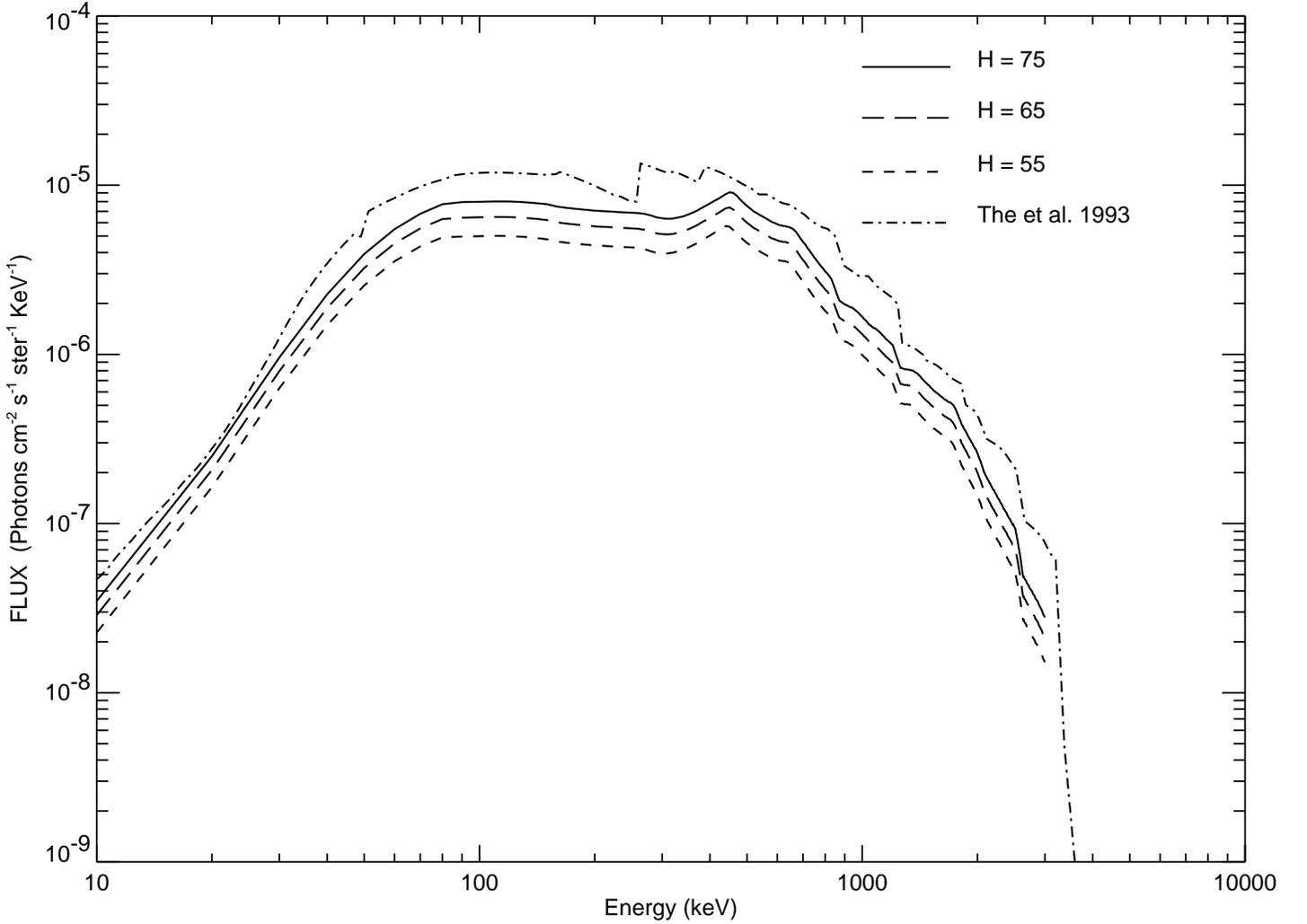}{300pt}{90}{80}{80}{324}{1}
\caption{The CGB contribution from SNI based on the SFR evolution function 
$\eta$ (eq. 12) with A = 4, B = 0.25, $z_p$ = 1, and 1.5 Gy time delay for SNI.
Three values for the Hubble constant are presented. For comparison we also show
the results of \cite{THE93}, multiplied by 2/3 (based on the assumption that 
2/3 of the cosmic iron production can be attributed to Type Ia supernovae; 
\cite{THIE91}). 
\label{cgbsnia}}
\end{figure}

We first present the background due to SNI in Figure~\ref{cgbsnia} for a few 
choices of the Hubble constant (h= 0.55, 0.65, and 0.75). It is clear that 
supernovae can not contribute to the CGB above 3.5 MeV, simply because they
do not produce radioactive isotopes that emit $\gamma$-ray lines above that 
energy. The presence of strong lines leads to steps in the spectrum that might 
be observable with future $\gamma$-ray instruments. If one had a dominant 
strong single line, the slope of the cosmologically broadened spectrum would 
be a direct measure of the cosmic star formation history. In reality the 
line(s) are blended with the Compton continuum, and the detailed shape of the 
spectrum can not be resolved with current technology. Still, if it is true that
the CGB in the MeV regime is $\sim$ 100$\%$ due to emission from supernovae, 
then the spectrum shown in Figure~\ref{cgbsnia} provides an independent and 
unique measure of the universal star formation history! For comparison we also 
display previous results derived by \cite{THE93}, who used a simple model of Fe 
synthesis in the universe to derive the CGB. We scale their result by a factor 
2/3, in order to take into account estimates of the fraction of iron that is 
contributed globally by supernovae of Type Ia (\cite{THIE91}). 

\placefigure{cgbsnia}

\begin{figure}
\plotfiddle{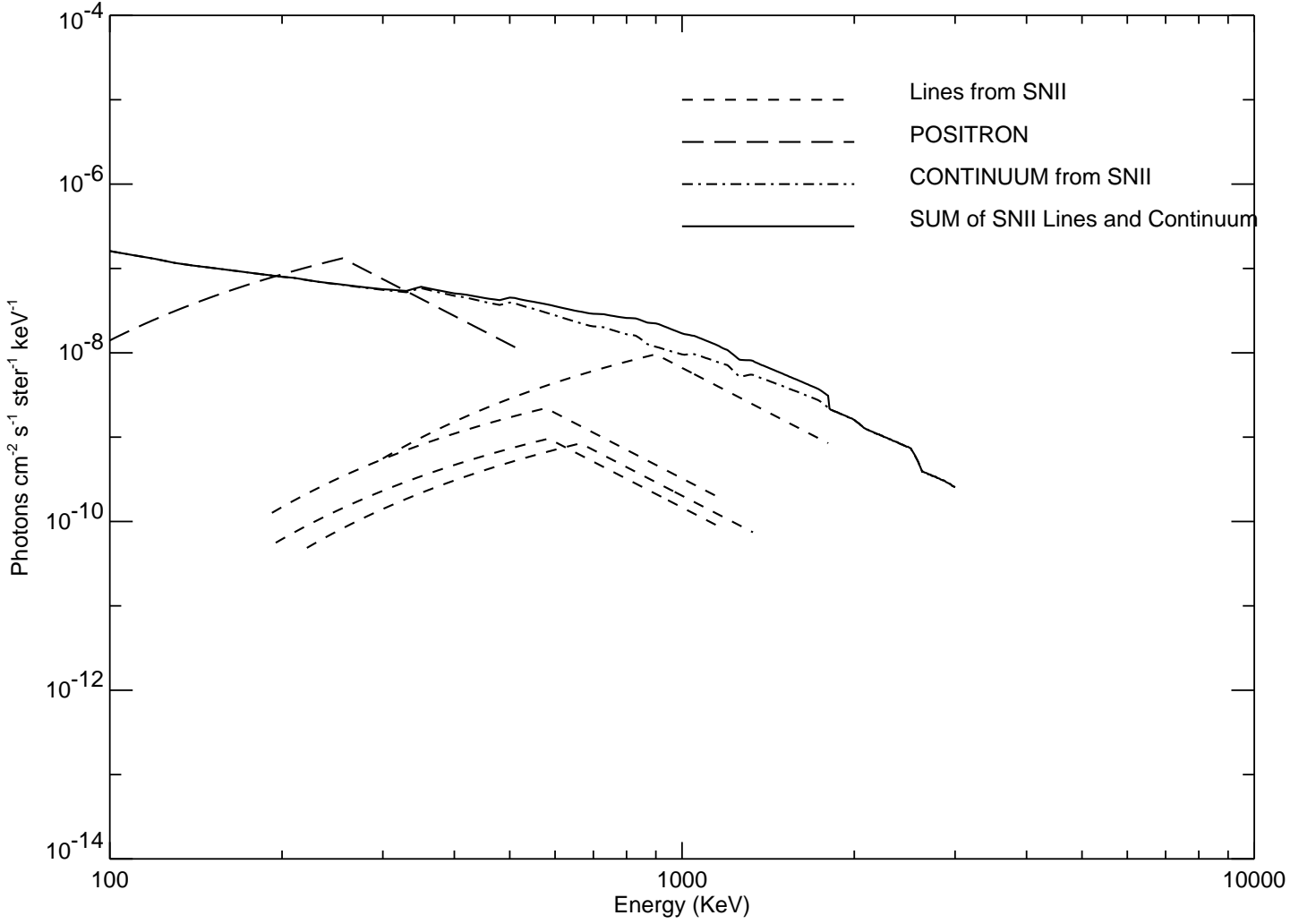}{300pt}{90}{80}{80}{324}{1}
\caption{Type II supernovae contributions to the CGB from \(^{56}{\rm Ni} 
\rightarrow\)  \(^{56}{\rm Co} \rightarrow\) \(^{56}{\rm Fe} \) decay 
$\gamma$-lines and their corresponding comptonized $\gamma$-ray photons 
including long-lived line afterglows due to $^{26}$Al, $^{44}$Ti, $^{60}$Co, 
and positrons. 
\label{cgbsnii}}
\end{figure}

Next we show the contributions to the CGB from Type II supernovae 
(Figure~\ref{cgbsnii}), and include contributions from long-lived isotopes 
($^{26}$Al, $^{44}$Ti, $^{60}$Co) as well as positrons. 
It is clear that these 
nuclei do not contribute substantially to the MeV background. The SNII 
contribution is small compared to that from SNI. Integrating the spectra, we 
estimate that the II/I ratio is $\sim$ 0.01, as also pointed out by 
\cite{THE93}.  

\placefigure{cgbsnii}

\begin{figure}
\plotfiddle{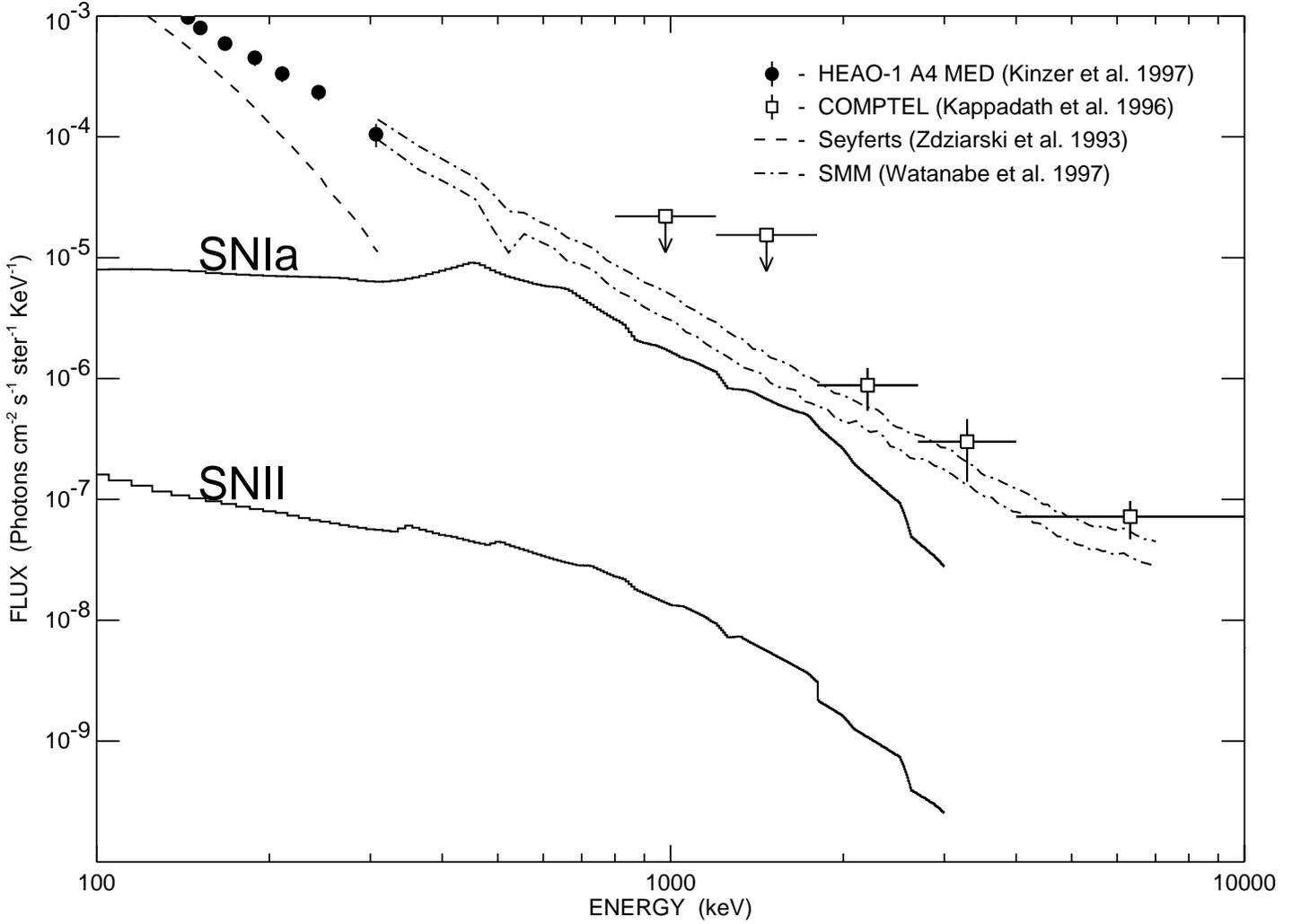}{300pt}{90}{80}{80}{324}{1}
\caption{The CGB from SNI, SNII (including long-lived isotopes),
in comparison to recent CGB measurements from
HEAO-1 A4 (\cite{KINZ97}), COMPTEL (\cite{KAPP96}), 
and SMM (\cite{WATA97}). For the SMM data the 
dashed-dotted lines indicate the $\pm$ 1$\sigma$ region of uncertainty
of the averaged spectrum. Predictions for the background contributions
from Seyfert galaxies (\cite{ZDZI93}) are also shown (dashed line).
\label{all2}}
\end{figure}

In Figure~\ref{all2} we plot the theoretical estimates together with existing
measurements of the CGB. The {\it standard} model is apparently able to
explain the bulk of the $\sim$ 1 MeV observed flux. In fact, for certain 
choices of model parameters $\gamma$-ray overproduction occurs, so that we are 
in principle able to constrain cosmological parameters and the cosmic star 
formation history. From the comparison of the different contributions it is 
clear that the MeV background is dominated by $\gamma$-rays from SNI. However, 
not all questions are answered. Even if the CGB is due to SNI, we must explain 
the apparent absence of a sharp drop in the observed flux above 3.5 MeV, where
supernovae do not emit at all. Is it reasonable to expect blazars to fill in
the gap in just the right way to explain the nearly continuous power-law
behavior of the CGB? Too little is known about blazar spectra in this energy
range (\cite{BLOE95}; \cite{BLOM95}). The smooth continuation of the SMM data
from 2 MeV to 6 MeV might indicate that SNI contribution to the flux near 1 MeV
is only a small fraction. A similar problem is apparent at lower energy (few 
100 keV), where Seyfert spectra fall rapidly and supernovae create only a flat 
spectrum.
 
\placefigure{all2}

Let us place this result (Figure~\ref{all2}) in perspective: the most
plausible origin of the CGB in the soft $\gamma$-ray region ($< 0.5$ MeV) 
appears to be a sum of unresolved Seyfert galaxies (e.g. \cite{ZDZI93};
\cite{ZDZI96}), which is consistent with balloon observations (e.g. 
\cite{KINZ78}). In the hard $\gamma$-ray region ($> 10$ MeV), unresolved 
blazars are successfully invoked (e.g. \cite{STEC96}) to match high-energy 
EGRET observations (\cite{CM98}; \cite{SREE98}). The spectral properties of 
Seyfert galaxies and blazars leave a window around 1 MeV in which their 
integrated fluxes would be far below the observed level (\cite{KAPP96}; 
\cite{WATA97}). It is perhaps reassuring that our standard model for the 
supernova contribution is within a factor two of the observations, which argues 
for SNI as the primary explanation of the bulk, or all, of the MeV background.
This fact is exploited below to generate constraints on the cosmic star 
formation history. However, we note the discrepancy between the combined 
Seyfert/Ia prediction and the data around 400 keV, perhaps hinting at a so far 
undetected population of sources or else suggesting that a fraction of Seyferts
may have harder spectra than those found so far with the Compton Observatory. 

As apparent in Figure~\ref{all2} the line features
at 847, 1238, 1770, 2030 and 2599 {\rm keV} due to $\gamma$-ray line emission 
from the decay chain  \(^{56}{\rm Ni} \rightarrow \)  \(^{56}{\rm Co} 
\rightarrow\) \(^{56}{\rm Fe} \) are not washed out completely, but still 
imprint a step structure in the CGB spectrum. So far one (unsuccessful) attempt 
was made to detect these edges (\cite{BART96}), leaving a significant  
challenge for other current and also for future observations in the MeV regime.

\subsection{Progenitors of Type Ia Supernovae}

\begin{figure}
\plotfiddle{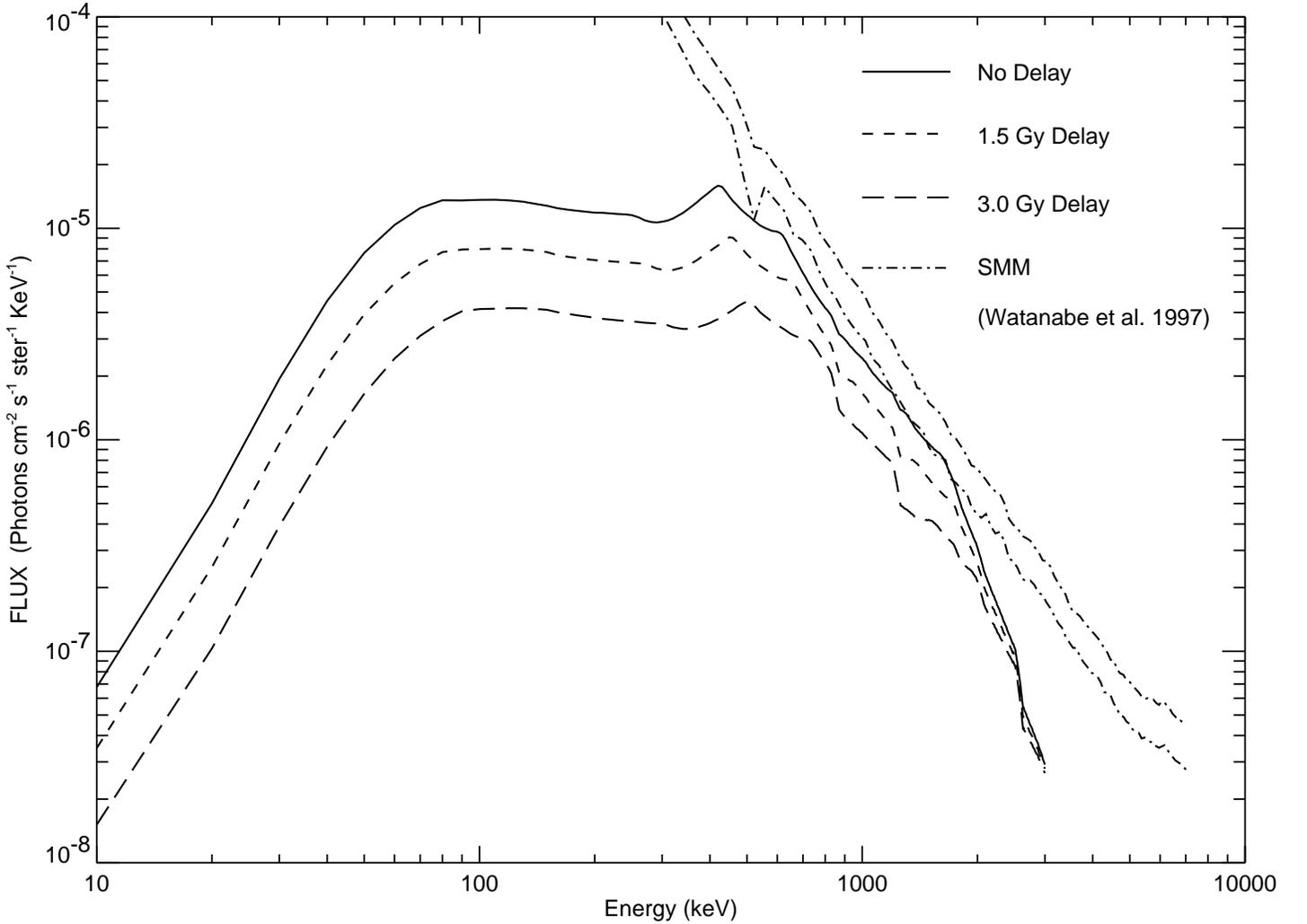}{300pt}{90}{80}{80}{324}{1}
\caption{Effects of the characteristic explosion/delay time scale 
($\tau_I$, see text) of SNI relative to SNII which directly trace the cosmic
star formation history. Three cases are shown:$\tau_I$ = 0.0, 1.5, and 
3.0 Gy. Note that we keep the SNI/SNII ratio at z = 0 fixed at 1/3 in all
cases. The comparison to the preliminary SMM estimates of the CGB 
(\cite{WATA97}) shows that large delays lead to a CGB flux that significantly 
falls short of the observed spectrum.
\label{delay}}
\end{figure}

A satisfactory progenitor scenario for SNI has not yet emerged (e.g., 
\cite{NIK97}; \cite{TM98}). Leading models invoke single-degenerate (SD) 
(\cite{WHEL73}; \cite{IBEN84}) or double-degenerate (DD) systems (\cite{WEB84};
 \cite{IBEN84}),i.e., distinguished by the nature of the white dwarf companion. 
Models of the explosion are sensitive the mass of the white dwarf. One 
refers to Ch or sub-Ch models, depending on the mass of the unstable white 
dwarf relative to the Chandrasekhar limit. We might learn something about the 
right progenitor systems through observations of the delay between star 
formation and Ia occurrence, which one might be able to derive from the 
observation of the relative I/II supernova rate as a function of redshift 
(\cite{MADA98}; \cite{SADA98}; \cite{RUIZ98}). Timescales of SNI explosion in 
both model classes depend on various parameters, such as the initial separation
and the mass ratio of two WDs in DD models and length of time spent filling the
Roche lobe in SD models (\cite{MADA98}). This time delay of SNI explosion 
after the global SFR can be $\sim 10^{9}$ yrs or as short as $\sim 10^{7}$ yrs
(e.g., \cite{RUIZ98}), depending on the particular scenario under consideration.
Following \cite{MADA98} we employed a characteristic explosion (delay) time 
scale ($\tau_I$), defining the explosion probability per white dwarf, and 
assume $\tau_I$ to be epoch independent. Given the vast uncertainties in the Ia
heritage, we consider three cases:$\tau_I$ = 0.0, 1.5, and 3.0 Gy. 
Figure~\ref{delay} shows the effects of these choices. Note that we keep the 
ratio of I/II = 1/3 at z=0 for any cases. 

\placefigure{delay}

\subsection{Constraints on the Cosmic Star Formation History}

\begin{figure}
\plotfiddle{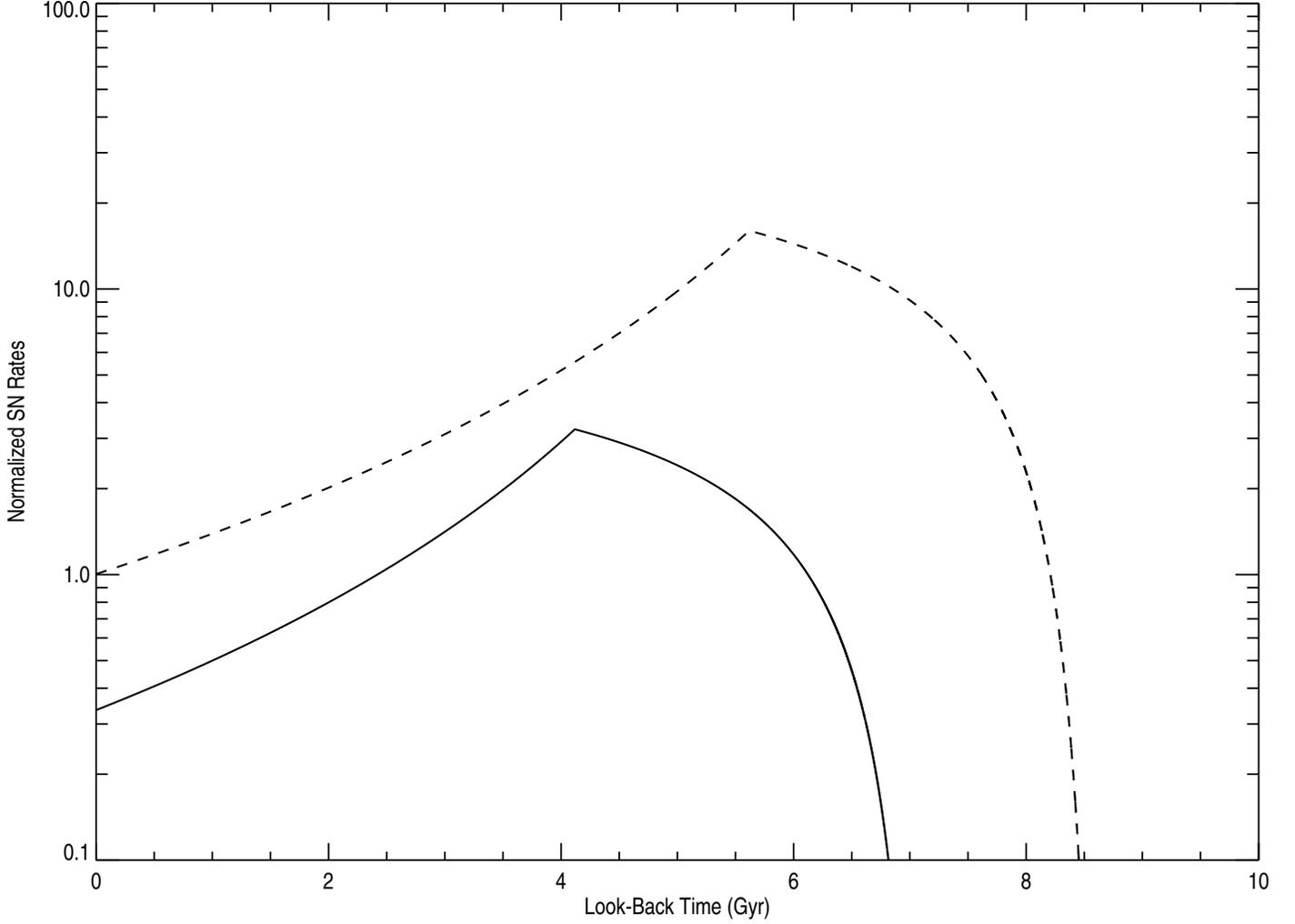}{300pt}{90}{80}{80}{324}{1}
\caption{Normalized SN rates in function of look-back time. The SNII rate 
(dashed line) is normalized at today. The SNI rate (solid line) has a delay
time from a birth of a WD to the SN explosion. For simplicity we adopted a 
mean lifetime of SNI progenitors of 1.5 Gy which was obtained from the 
chemical evolution study (\cite{YOSH96}). Note that different $\xi_{SN}$ in Eq. 
(6) are used for each type of SN.  
\label{snr_look_back}}
\end{figure}

\begin{figure}
\plotfiddle{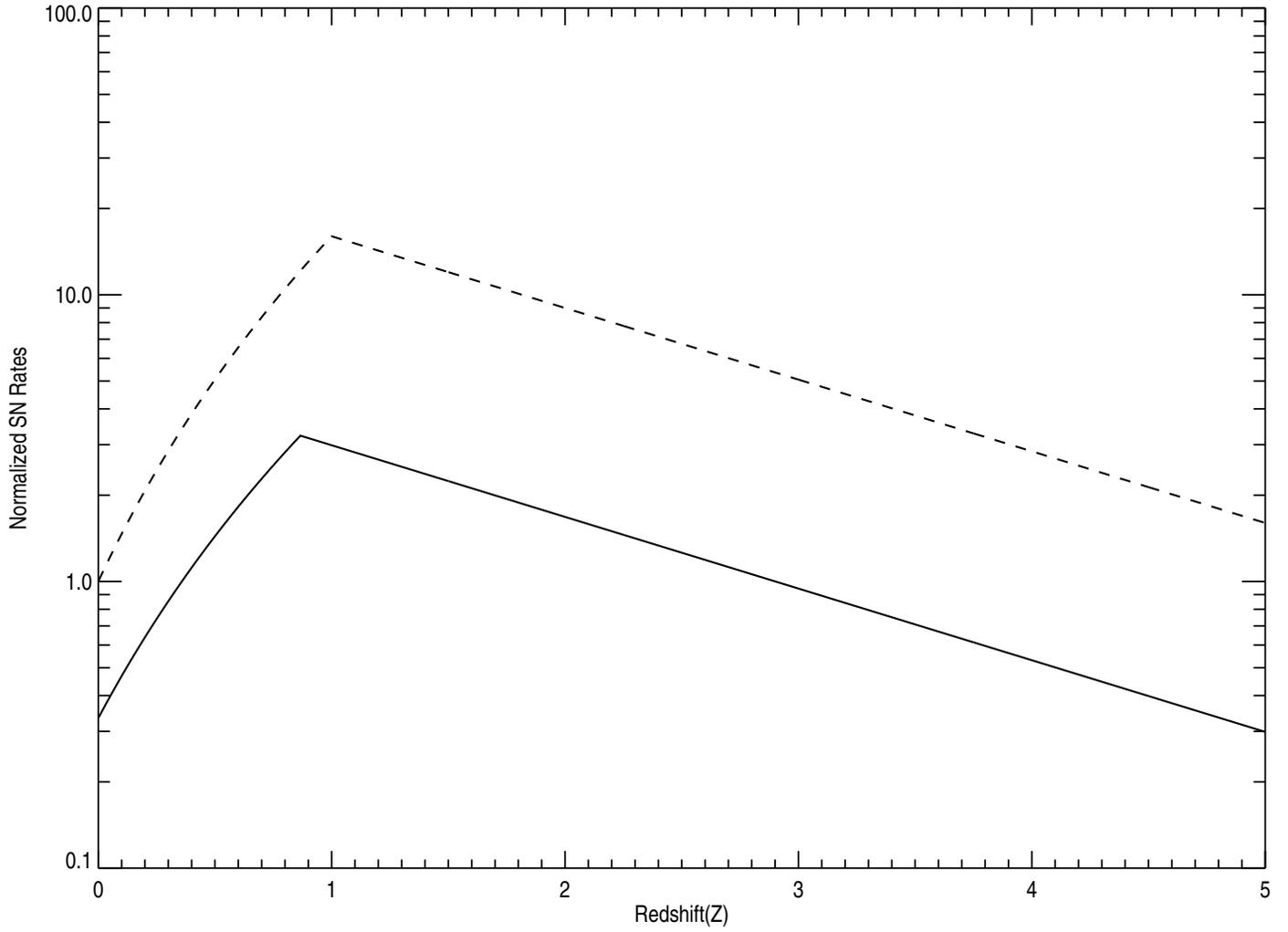}{300pt}{90}{80}{80}{324}{1}
\caption{Same as Figure~\ref{snr_look_back}, but as a function of redshift (z).
Transformation between z and t involves a model dependent look-back time.
\label{snr_z}}
\end{figure}

\begin{figure}
\plotfiddle{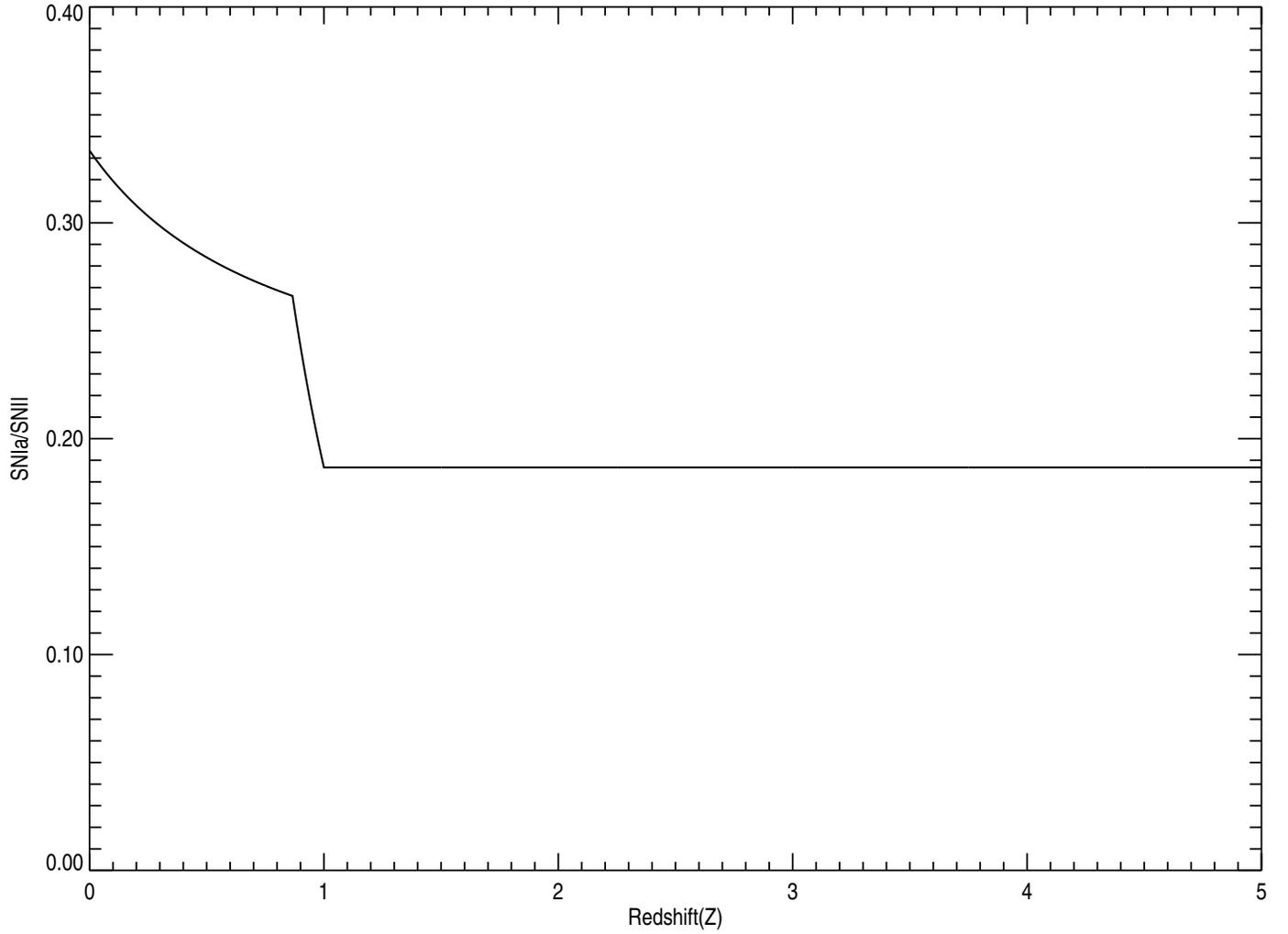}{300pt}{90}{80}{80}{324}{1}
\caption{The ratio of the SNI rate to the SNII rate as a function of redshift. 
The ratio is nearly constant for $z > 1$ but rapidly increases towards 
small z, reaching 1/3 at z = 0 according to our adopted normalization for the
present-day derived from local supernovae surveys (see text). \label{sn_ratio}}
\end{figure}

Figure~\ref{snr_look_back} shows the evolution of the supernova rate as a 
function of look-back time (which depends somewhat on our choice of 
cosmological parameters). Figure~\ref{snr_z} shows the same functions, but 
displayed in redshift space. The normalization is $R_{II}$(z=0) = 1. As 
discussed above, the function for SNII is identical to (our fit of) the observed
star formation rate history, while the SNI curve shows a delay, but is 
normalized to give a present-day ratio of SNI/SNII = 1/3. This ratio 
decreases with redshift (Figure~\ref{sn_ratio}), resembling the trend predicted
by population synthesis models (e.g., see Figure 5 in \cite{YUNG98}). 

\begin{figure}
\plotfiddle{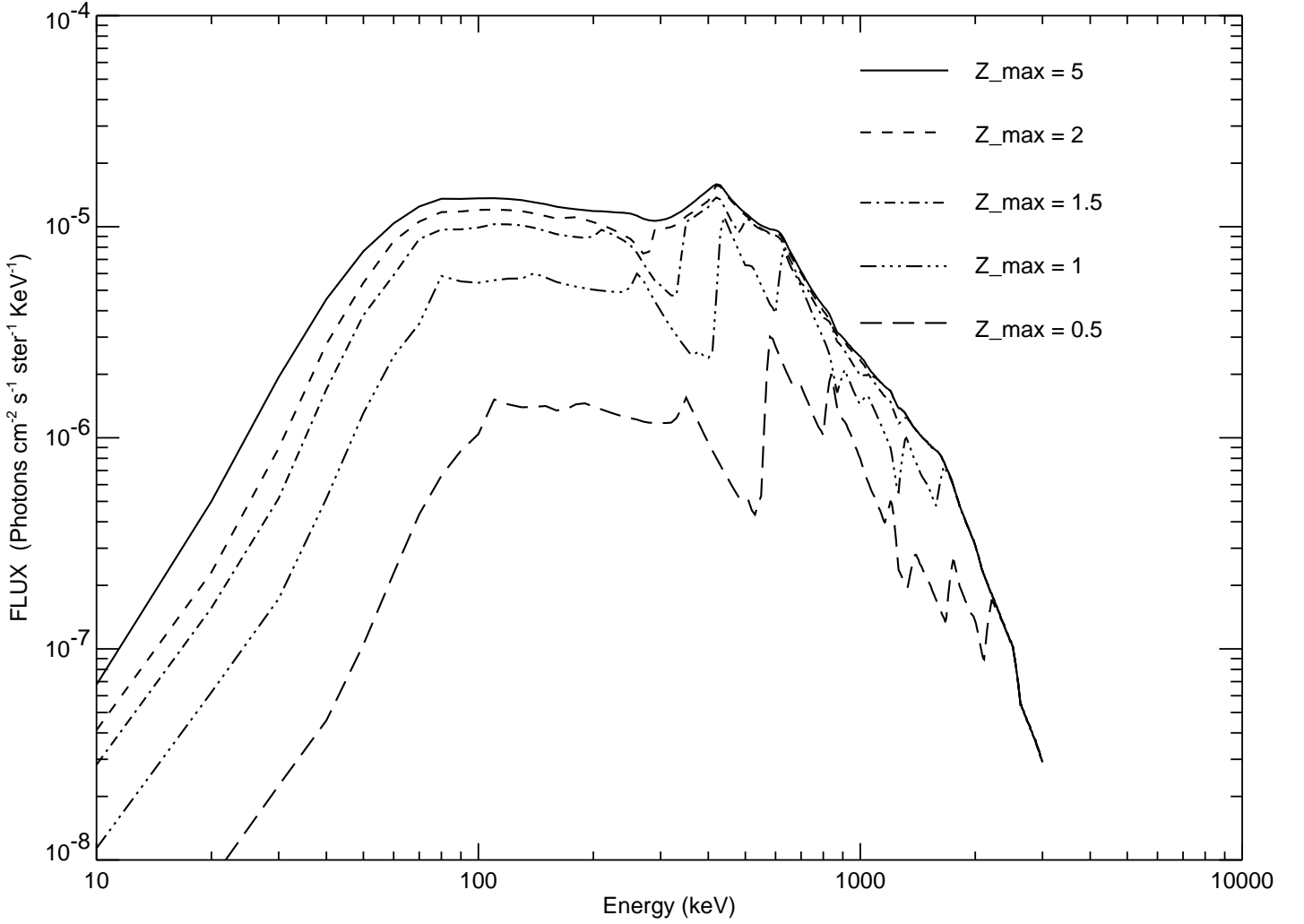}{300pt}{90}{80}{80}{324}{1}
\caption{Curves indicating how the CGB grows with integration to maximum 
redshifts 0.5, 1.0, 1.5, 2.0 and 5.0 (no SNI delay was assumed for this 
calculation). The figure illustrates that most of the CGB is due to emission 
from redshifts smaller than the peak resdhift of the cosmic star formation 
history.
\label{z_max}}
\end{figure}

Figure~\ref{z_max} shows multiple curves indicating how the CGB grows with 
maximum redshift included. Here redshifts up to 0.5, 1.0, 1.5, 2.0 and 5.0 are
plotted, showing that most of the CGB comes from SFR at redshift less than 
$\sim$ 1.5, so that our results are insensitive to the possibly large 
uncertainties in measurements of the cosmic star formation rate at high 
redshifts. We also considered star formation histories in which the rate 
remains constant between z =  z$_p$ and z = 5, but this affects only emission 
from high redshift so that the correspondingly shifted supernova spectra 
contribute in an energy regime that is completely dominated by other sources. 
In other words, measurements of the MeV background are not able to constrain 
the SFR at high redshift. The situation is far better for the recent SFR 
history, which is the last point of discussion.  

\placefigure{snr_look_back}
\placefigure{snr_z}
\placefigure{sn_ratio}
\placefigure{z_max}

For a given delay time $\tau$ the supernova-induced CGB is mostly determined by 
the location and strength of the peak in the cosmic star formation history. In
general, a shorter delay and/or smaller $z_p$ will increase the CGB. If the 
delay is zero, the CGB is inversely proportional to the location of $z_p$ (in 
our simple representation of the SFR with two power laws). Placing the time of 
the peak of star formation rate at larger redshift reduces the CGB. Of course, 
the total integrated flux also depends on the assumed I/II ratio, and the 
overall amplitude of the SFR. The current observational status of the SFR 
history (e.g. \cite{LILL96}; \cite{CONN97}; \cite{SADA98}; \cite{MADA98}; 
\cite{Guzman98}) suggests values around (z$_p$, $\eta(z_p)$ = (1$-$2,10$-$30). 
These values do not lead to a conflict with the measured CGB. 
Any increase in nickel yields in an average SNI or a global increase in the 
I/II ratio would eventually conflict with the CGB observations, which therefore
provides an independent constraint on the global properties and 
cosmic evolution of Type Ia supernovae. While the constraints on cosmological
parameters (H$_0$ and $\Omega$) do not lead to improvements over other methods,
the CGB provides a unique handle on the unknown delay parameter $\tau_I$.

While the increase of the SFR with redshift by a factor 20$\pm$10 to the peak
location z$_p$ remains a robust conclusion of recent studies of the SFR, the
normalization has been challenged. A new determination of the local
volume-averaged star formation rate from the 1.4 Ghz luminosity function of star
forming galaxies (\cite{SGO98}) implies a local SFR density 2$-$3x larger than 
the Gallego {\it et al.} H$_\alpha$ estimate. Tresse and Maddox (1998) have
shown the H$_\alpha$ luminosity function at z = 0.2, and their data suggest
a SFR twice that determined from UV measurements. A similar shortfall of
UV-based estimates in comparison to those derived from Balmer lines was
reported by \cite{GLA98} who used J-band IR spectroscopy of a redshift
selected sample of 13 galaxies from the Canada-French Redshift Survey (CFRS)
to measure the H$_\alpha$ luminosity function at z = 1. These observations
also indicate elevated SFR values (factor 2$-$3) relative to values derived
from the UV. We already mentioned the deep submillimeter survey of the HDF
(\cite{HUGH98}), which implies a star formation rate for z = 2$-$4 that is
five times higher than that derived from optical and UV observations of the 
HDF. The interpretation of these differences involves the well known fact
that star formation occurs in dense molecular regions which hides the optical
and UV emission due to substantial extinction. That much of the cosmic star
formation activity occurs in very dusty regions is also supported by 
observations of the IR background in the 140$-$240 $\mu$m region, which was 
recently detected by the DIRBE and FIRAS experiments aboard COBE 
(\cite{DWEK98} and references therein). The IR background data also show that 
the UV and optically determined star formation rates fall short in producing 
the IR background, and specifically require the peak star formation rate (at z 
$\sim$ 1.5) to be larger by at least a factor two (\cite{DWEK98}). These 
various recent claims strongly suggest that the SFR(z) function we use as 
``standard model'' should be multiplied by a factor 2$-$3 {\it at all 
redshifts}! This would simply mean that we have to multiply the CGB fluxes by 
the same factor. Without the delay of SNI explosions this would yield a CGB 
spectrum in excess of the observed values, while for a 3 Gyr delay the model 
matches the observed flux. With the revised SFR values we can thus explain all 
of the CGB with emissions from Type Ia supernovae, and one does not need to 
invoke a new source population. However, a delay of 3 Gy is on the extreme 
side of the suggested values. It is clear that the CGB significantly 
constrains the properties of SNI and possible further increases in SFR values.
If future observations can provide an accurate functional form for SFR(z), the 
CGB can be used to constrain supernova models. However, if SFR(z) estimates 
continue to be revised, the CGB provides a useful upper limit.

\section{Conclusions}

We calculated the contribution of supernovae to the cosmic $\gamma$-ray 
background (CGB). Following \cite{THE93} we used models W7fm and W10HMM as
source templates for SNI and SNII, respectively. Our approach differs
from that of \cite{THE93} through the use of the observed star formation 
history of the universe, obtained from Ly$\alpha$ QSO absorption studies (e.g.,
\cite{PEI95}), galaxy redshift surveys (e.g., \cite{LILL96}), broad band 
photometry of galaxies in the Hubble Deep Field (e.g., \cite{MADA96}),and other
recent work in this very active area of observational cosmology. We consider 
time delays of 0$-$3 Gy between SNII and SNI explosions. Our 
estimated background spectra are similar to those derived by \cite{THE93}. We 
confirm their finding that SNII contribute very little ($\sim$ 1 \%) to the 
CGB. SNI on the other hand, could explain most or all of the observed CGB 
spectra, for certain parameter choices. However, the standard model does not
match the observed flux of the CGB, which suggests that either the current flux 
measurements are still an overestimate, that there could still be an 
unrecognized source population making a substantial contribution to the MeV 
background, or that the cosmic star formation rate is significantly higher than
commonly assumed. This discrepancy would become even more serious if the SNI 
delays were much larger than 1 Gy, and/or if most of the cosmic star formation 
activity occurred at redshifts past z = 1. For example, combining Madau's rates 
and a 2 Gy delay leads to a CGB flux that falls short of the observations by 
more than a factor three. On the other hand, short delays (less than 1 Gy) 
combined with a star formation history that has an increase of 10-30 by z$_p$ =
1 yield a CGB flux that is just sufficient to explain the observations in a 
limited range of photon energies. We conclude from this, that the currently 
favored scenarios of SNI progenitors and their cosmic rate evolution 
underpredict the CGB. Recent upward revisions of the cosmic star formation 
history at all redshifts increase the predicted CGB fluxes, compensating the 
shortfall. However, an increase by a factor 2$-$3 overproduces the CGB unless 
the SNI delay time scale is much larger than 1 Gyr.

While it can not be proven, it also can not be ruled out that SNI could provide 
most or all of the CGB. If that were the case, theory predicts strong spectral 
steps in CGB due to line features from the \(^{56}{\rm Ni} \rightarrow \)  
\(^{56}{\rm Co} \rightarrow\) \(^{56}{\rm Fe} \) decay. These have not yet been
observed, but remain a challenge to current and future $\gamma$-ray experiments.
Some fraction of the CGB (mostly below each of the major line energies) must be 
non-isotropic, because the high energy end of each line is due to emission from
the nearest galaxies, which are known to have a very non-uniform spatial 
distribution to distance of $\sim$ few 100 Mpc (or redshift of z $\sim$ 0.1). 
The angle averaged spectrum out to z = 0.1 integrated over energy (thus the 
number of photons $cm^{-2} s^{-1}ster^{-1}$) is about 2\% of the total 
(integrated to z = 5). Thus, to see the anisotropy in this energy range one 
would need new detectors that can detect the CGB to 1\% accuracy.

The $\gamma$-ray background in the MeV regime provides valuable constraints on 
global iron synthesis in the universe, the global rate of star formation, and 
yields and lifetimes of SNI progenitors. These constraints are complimentary to
other astronomical methods. Current observations of the CGB in the MeV range 
can largely be explained as the unresolved superposition of $\gamma$-rays 
emitted by Type Ia supernovae. At present there is no unsurmountable conflict 
between the estimated SFR, I/II supernova rate ratio, nickel yields, Ia 
explosion time scales, and the observed CGB flux. However, the parameter ranges
are significantly constrained and there are open questions about the quality 
of the CGB fit from the combination of Seyfert galaxies, SNe, and blazars.  


\clearpage

\clearpage

\clearpage

\clearpage

\clearpage

\clearpage

\clearpage

\clearpage

\clearpage

\end{document}